# Role of higher-order effects in spin-misalignment small-angle neutron scattering of high-pressure torsion nickel

Y. Oba[1,2*], M. Bersweiler[2], I. Titov[2], N. Adachi[3], Y. Todaka[3], E.P. Gilbert[4], N.-J. Steinke[5], K.L. Metlov[6], A. Michels[2]

[1]*Materials Sciences Research Center, Japan Atomic Energy Agency, 2-4 Shirakata, Tokai, Ibaraki, 319-1195, Japan*

[2]*Department of Physics and Materials Science, University of Luxembourg, 162A Avenue de la Faïencerie, L-1511 Luxembourg, Grand Duchy of Luxembourg*

[3]*Department of Mechanical Engineering, Toyohashi University of Technology, 1-1 Hibarigaoka, Tempaku, Toyohashi, Aichi 441-8580, Japan*

[4]*Australian Centre for Neutron Scattering, Australian Nuclear Science and Technology Organisation, Locked Bag 2001, Kirrawee DC, NSW 2232, Australia*

[5]*Institute Laue-Langevin, 71 avenue des Martyrs, F-38042 Grenoble, France*

[6]*Donetsk Institute for Physics and Technology, R. Luxembourg str. 72, 83114 Donetsk, Ukraine and Institute for Numerical Mathematics RAS, 8 Gubkina str., 119991 Moscow GSP-1, Russia*

**ABSTRACT**

Magnetic-field-dependent unpolarized small-angle neutron scattering (SANS) experiments demonstrate that high-pressure torsion (HPT) straining induces spin misalignments in pure Ni, which persist in magnetic fields up to 4 T. The spin-misalignment scattering patterns are elongated perpendicular to the applied magnetic field due to an unusual predominant longitudinal $\sin^2\theta$-type angular anisotropy. Such a contribution cannot be explained by the conventional second order (in spin misalignment amplitude) micromagnetic SANS theory in the approach-to-saturation regime, nor can its magnitude relative to the other features of the cross sections by the third order micromagnetic SANS theory. This indicates that the high-density of crystal defects induced via HPT straining in Ni makes such higher-order effects in the micromagnetic SANS cross sections observable.

**I. INTRODUCTION**

High-pressure torsion (HPT) is one of the well-known severe plastic deformation techniques used for the synthesis of ultrafine-grained and nanocrystalline materials [1-5]. This method attracts attention because it drastically changes the microstructure of materials by significant grain refinement and by inducing a high-density of crystal defects. While it is well known in metallurgy that HPT microstructures can exhibit improved mechanical properties such as tensile strength, the HPT process also gives rise to fascinating effects

---

in magnetism. For example, previous studies have reported the formation of novel magnetic materials [6,7] and the onset of ferromagnetism in otherwise nonmagnetic elements [8,9]. The characteristic microstructures produced by HPT can also affect the magnetic properties via the pinning of magnetic domain walls at defect sites, changes in magnetocrystalline anisotropy by the distortion of crystallographic symmetries, or via the modification of the exchange coupling between the refined grains [10-12]; however, such effects of HPT on the magnetic properties still need to be further clarified.

Several researchers have approached this issue using pure Ni processed via HPT straining (HPT-Ni) [13-17]. Nickel is one of the simplest ferromagnets at room temperature with a stable face-centered cubic crystal structure and suitable as a model system to investigate the fundamentals of the effects induced via HPT straining. In addition, the change in the microstructure in Ni often brings about unique magnetic properties, *e.g.*, a lower Curie temperature and a smaller saturation magnetization in amorphous Ni and enhanced magnetic anisotropy in nanocrystalline Ni [18,19]. Scheriau *et al.* and Mulyukov *et al.* reported an increase of the coercivity as well as a decrease of the saturation magnetization in HPT-Ni [13,14,17]. These results were interpreted as possible effects of the crystal defects. On the other hand, Lorentz electron microscopy confirmed the presence of large magnetic domains, expanding over a large number of grains in HPT-Ni [15,16]. Although the magnetic domain structure was chaotically composed of domain walls passing along grain boundaries and across grains, no difference was observed between HPT-Ni and coarse-grained Ni.

Previous investigations suggest that the effect of defects on the spin texture may be the key to understand the effects of HPT straining on the magnetic properties. Here, we employ magnetic-field-dependent unpolarized small-angle neutron scattering (SANS) to clarify the role played by the defects in HPT-Ni. SANS is a powerful technique to characterize, on the mesoscopic length scale (~1-500 nm), the crystalline as well as the magnetic microstructure of bulk magnetic materials [19-26]. In a very recent unpolarized SANS study on HPT-Ni [26], we have analyzed the field dependence of real-space magnetic correlations and found that the characteristic sizes of the spin disorder vary on a scale between about 10-30 nm. From the field-variation of the magnetic correlation length, the defect size and an increased effective magnetic anisotropy (relative to the single-crystalline reference state) could be estimated. While the data analysis in [26] is based on the model-independent computation of the magnetic correlation function, in the present paper, we focus on the analysis of the angular anisotropy of the two-dimensional scattering pattern. We report on the observation of an unusual $\sin^2\theta$-type angular anisotropy of the *spin-misalignment SANS cross section* (following subtraction of nuclear and magnetic SANS at saturation) in the approach-to-saturation regime. The magnetic SANS data analysis and the discussion in the present paper is based on the micromagnetic SANS theory developed in Refs. [20-23].

## II. EXPERIMENT

A sheet of Ni was cut into disks with a diameter of 10 or 20 mm. The HPT process was conducted under standard conditions at a pressure of 5 GPa, with a rotation number of 5, and a rotational speed of 0.2 rpm [3]. After the HPT process, both sides of the disks were mechanically polished to adjust the sample thickness to 0.5 mm. Disks of non-deformed (nd) Ni with the same dimensions were also prepared for comparison.

Magnetization curves were measured using a Cryogenic Ltd. vibrating sample magnetometer (VSM) equipped with a 14 T superconducting magnet.

The SANS experiments were carried out using the QUOKKA instrument of the Australian Nuclear Science and Technology Organisation (ANSTO) [27]. Four 20 mm disks were stacked for measurements. A superconducting magnet was used to apply a magnetic field between 1 and 10 T to the samples. The direction of the applied magnetic field **H** was perpendicular to the incident neutron beam and parallel to the disk surface, which corresponds to the so-called perpendicular scattering geometry (compare Fig. 1) [21]. The measurements were performed using equal source-to-sample and sample-to-detector distances of 20 m and 8 m. The average neutron wavelength $\lambda$ was set to 0.5 nm with 10 % wavelength resolution (FWHM). These conditions allowed one to measure a $q$-range from 0.03 nm$^{-1}$ to 1.0 nm$^{-1}$, where $q$ denotes the magnitude of the scattering vector, which for elastic scattering is given by $q = 4\pi/\lambda \sin(\psi/2)$ with $\psi$ being the scattering angle. The obtained scattering patterns were corrected for background scattering, sample transmission and detector sensitivity [28].

Additional SANS experiments under similar conditions were performed at the instrument D33 at the Institut Laue-Langevin (ILL) [29]. Two disks of 10 mm diameter were stacked. The applied magnetic field to the sample was varied between 0.1 T and 6.7 T. In addition to the perpendicular scattering geometry, SANS measurements were also conducted in the so-called parallel scattering geometry, where the direction of the applied magnetic field is parallel to the incident neutron beam and normal to the disk surface (Fig. 1) [21]. The values of $\lambda$ and $\Delta\lambda/\lambda$ were 0.46 nm and 10 %, respectively. The samples were held at room temperature in all the SANS measurements. The D33 data were corrected for background scattering, sample transmission, and detector efficiency using the GRASP software package [30].

## III. MICROMAGNETIC SMALL-ANGLE NEUTRON SCATTERING THEORY

Based on micromagnetic SANS theory, the unpolarized elastic SANS cross section in the perpendicular scattering geometry can be written as [21-23]:

$$\frac{d\Sigma}{d\Omega}(\mathbf{q}, H) = \frac{d\Sigma}{d\Omega_{res}}(\mathbf{q}) + \frac{d\Sigma}{d\Omega_{SM,\perp}}(\mathbf{q}, H), \tag{1}$$

where

$$\frac{d\Sigma}{d\Omega_{res}}(\mathbf{q}) = \frac{8\pi^3}{V}\left(|\widetilde{N}(\mathbf{q})|^2 + b_H^2|\widetilde{M}_s(\mathbf{q})|^2 \sin^2\theta\right) \tag{2}$$

represents the nuclear and magnetic residual scattering contribution, which is measured at complete magnetic saturation. In Eqs. (1) and (2), $V$ is the sample volume, $b_H = 2.70 \times 10^{-15}$ m/$\mu_B$ ($\mu_B$: Bohr magneton), $\widetilde{N}(\mathbf{q})$ denotes the Fourier transform of the nuclear scattering length density function $N(\mathbf{r})$, $\widetilde{M}_s(\mathbf{q})$ is the Fourier transform of the spatially-dependent saturation magnetization $M_s(\mathbf{r})$, and $\theta$ is the azimuthal angle on the two-dimensional (2D) detector [$\theta = \angle(\mathbf{q}, \mathbf{H})$], where $\mathbf{q}$ and $\mathbf{H}$ are the scattering vector and applied magnetic field (compare Fig. 1). The magnetic field ($H$) dependence of the scattering intensity is included in $d\Sigma/d\Omega_{\text{SM},\perp}(\mathbf{q}, H)$, which corresponds to the purely magnetic SANS cross section due to transversal spin misalignment. This quantity can be expressed as:

$$\frac{d\Sigma}{d\Omega_{\text{SM},\perp}}(\mathbf{q}, H) = \frac{8\pi^3}{V} b_\text{H}^2 \left[ |\widetilde{M}_x|^2 + |\widetilde{M}_y|^2 \cos^2\theta + \Delta\widetilde{M}_z^2 \sin^2\theta - (\widetilde{M}_y \widetilde{M}_z^* + \widetilde{M}_y^* \widetilde{M}_z) \sin\theta \cos\theta \right]$$

$$= S_\text{H}(\mathbf{q}) R_{\text{H},\perp}(q, \theta, H) + S_\text{M}(\mathbf{q}) R_\text{M}(q, \theta, H) + \frac{8\pi^3}{V} b_\text{H}^2 \Delta\widetilde{M}_z^2(\mathbf{q}, H) \sin^2\theta, \quad (3)$$

where

$$\Delta\widetilde{M}_z^2(\mathbf{q}, H) = |\widetilde{M}_z(\mathbf{q}, H)|^2 - |\widetilde{M}_s(\mathbf{q})|^2, \quad (4)$$

and $\widetilde{\mathbf{M}} = (\widetilde{M}_x, \widetilde{M}_y, \widetilde{M}_z)$ is the Fourier transform of the magnetization vector field $\mathbf{M} = (M_x, M_y, M_z)$. Note that $\widetilde{M}_z \cong \widetilde{M}_s$ in the approach-to-saturation regime, so that the third term on the right side of Eq. (3) vanishes and $\widetilde{M}_z$ is regarded as field-independent in this field range [31,32]. In Eq. (3), $S_\text{H}(\mathbf{q}) R_{\text{H},\perp}(q,\theta,H)$ and $S_\text{M}(\mathbf{q}) R_\text{M}(q,\theta,H)$ denote the spin-misalignment scattering contributions due to perturbing magnetic anisotropy fields and magnetostatic fields, respectively. The former term is composed of the anisotropy field scattering function $S_\text{H}(\mathbf{q})$ and of the micromagnetic response function $R_{\text{H},\perp}(q,\theta,H)$ as follows:

$$S_\text{H}(\mathbf{q}) = \frac{8\pi^3}{V} b_\text{H}^2 |\widetilde{H}_p(\mathbf{q})|^2, \quad (5)$$

$$R_{\text{H},\perp}(q, \theta, H) = \frac{p^2}{2} \left[ 1 + \frac{\cos^2\theta}{(1 + p\sin^2\theta)^2} \right], \quad (6)$$

$$p(q, H) = \frac{M_s}{H[1 + l_\text{H}^2(H) q^2]}, \quad (7)$$

$$l_\text{H}(H) = \sqrt{\frac{2A}{\mu_0 M_s H}}, \quad (8)$$

where $\widetilde{H}_p(\mathbf{q})$ is the Fourier coefficient of the magnetic anisotropy field, $p(q,H)$ is a dimensionless function, and $l_\text{H}(H)$ denotes the exchange length of the field. The parameter $A$ is the exchange-stiffness constant and $M_s$ the macroscopic saturation magnetization of the sample. Similarly, $S_\text{M}(\mathbf{q}) R_\text{M}(q,\theta,H)$ is separated into the scattering function of the longitudinal magnetization $S_\text{M}(\mathbf{q})$ and into the corresponding micromagnetic response function $R_\text{M}(q,\theta,H)$; more specifically,

$$S_\text{M}(\mathbf{q}) = \frac{8\pi^3}{V} b_\text{H}^2 |\widetilde{M}_z(\mathbf{q})|^2, \quad (9)$$

$$R_{\mathrm{M}}(q,\theta,H) = \frac{p^2 \sin^2\theta \cos^4\theta}{\left(1 + p\sin^2\theta\right)^2} + \frac{2p\sin^2\theta \cos^2\theta}{1 + p\sin^2\theta}. \tag{10}$$

These expressions reveal that, in the approach-to-saturation regime, the behavior of the spin-misalignment SANS cross section is determined by the applied magnetic field, the magnetic anisotropy field, the magnetostatic field and the exchange field.

The spin-misalignment scattering for the parallel scattering geometry can be similarly decomposed as [21-23]:

$$\frac{d\Sigma}{d\Omega_{\mathrm{SM},\parallel}}(\mathbf{q},H) = S_{\mathrm{H}}(\mathbf{q})R_{\mathrm{H},\parallel}(q,H) + \frac{8\pi^3}{V}b_{\mathrm{H}}^2 \Delta\widetilde{M}_z^{\,2}(\mathbf{q},H), \tag{11}$$

$$R_{\mathrm{H},\parallel}(q,H) = \frac{p^2}{2}, \tag{12}$$

where $R_{\mathrm{H},\parallel}(q,H)$ denotes the micromagnetic response function of the perturbing magnetic anisotropy field. Since $R_{\mathrm{H},\parallel}(q,H)$ is independent of $\theta$ in this scattering geometry, the orientation distribution of the magnetic anisotropy field $S_{\mathrm{H}}(\mathbf{q})$ and of the function $\Delta\widetilde{M}_z^{\,2}(\mathbf{q})$ may be directly reflected in the azimuthal anisotropy of the scattering pattern.

**IV. RESULTS AND DISCUSSION**

Figure 2 shows the magnetization curves of HPT-Ni and nd-Ni normalized by the saturation magnetization of bulk Ni ($M_s$ = 55.5 Am$^2$/kg [33]). The value of $M_s$ in nd-Ni is in agreement with that of bulk Ni, whereas HPT-Ni has a 2.3 % smaller $M_s$. This is consistent with a previous study which reports that the decrease of $M_s$ is due to a grain-boundary phase with lower crystal symmetry [14].

The upper row in Fig. 3 shows the typical 2D total (nuclear + magnetic) SANS cross section of HPT-Ni measured in the perpendicular geometry at several applied magnetic fields $H$ (between 0.1 and 6.7 T). The patterns are (at all $H$) predominantly elongated perpendicular to the magnetic field and, as our analysis of the angular anisotropy reveals (see below), also show lobes of higher intensity roughly along the detector diagonals at the intermediate $H$ between 0.4 and 4 T. With increasing $H$ up to 4 T, the scattering intensity drastically decreases, whereas the perpendicular elongation and diagonal lobes remain. At larger fields, the scattering intensity changes only slightly. These results are significantly different from the scattering patterns of nd-Ni, which are isotropic and show no magnetic field dependence for applied fields of 1 T and 2 T (see Fig. 4). This implies that the SANS cross section of nd-Ni is, in that field regime, dominated by isotropic nuclear scattering and that the magnetic SANS cannot be resolved by the conventional SANS technique. Since the magnetic-field dependence of the total SANS cross section can result only from the spin-misalignment scattering contribution [compare Eqs. (1) – (4)], our SANS results indicate that the HPT process induces significant spin misalignment, which persists for applied fields between 0.1 T and 4 T. Despite the significant change in the SANS cross section, the magnetization of HPT-Ni is already 93.3 % and 99.7 % of $M_s$ at 0.1 T and 1 T, respectively. These results indicate that only a small fraction of the misaligned spins contributes to the field-dependent SANS cross section and the remaining majority of magnetic moments is in magnetic saturation. This is similar to previous theoretical and experimental results of the spin-misalignment scattering [21].

Since in general the nuclear SANS cross section is field-independent, we determined the spin-misalignment scattering contribution by subtracting the total SANS cross section at the highest available fields (6.7 T for the ILL data and 10 T for the ANSTO data) from the ones at lower fields. This specific neutron data analysis procedure has already been employed to extract the purely spin-misalignment SANS cross section in several other systems (e.g., [34,35]). The bottom row in Fig. 3 depicts the corresponding 2D magnetic SANS cross sections. The vertical elongation and the diagonal lobes remain after the subtraction procedure and thus are attributed to the spin-misalignment scattering. This is different from previous reports of magnetic SANS from nanostructured magnetic materials, where the spin-misalignment scattering contributions feature a flying-saucer-type scattering pattern [due to the term $R_{H,\perp}(q,\theta,H)$] and a clover-leaf-type pattern [due to $R_M(q,\theta,H)$] represented as Eqs. (6) and (10), respectively [36-39].

Figure 5 displays the magnetic SANS cross sections of HPT-Ni (radially-averaged over $2\pi$) at applied fields between 0.1 T and 4 T. The cross sections for fields higher than 4 T manifest only negligible intensities within experimental error. With increasing $H$ between 0.1 and 4 T, the scattering intensities monotonically decrease by more than three orders of magnitude at $q = 0.04$ nm$^{-1}$. The slope in the high-$q$ region is steeper than $q^{-4}$, which is a typical feature of the spin-misalignment scattering contribution (e.g., [21,26,34]). We also emphasize that the majority of the SANS measurements reported here fall into the approach-to-saturation regime (compare to the magnetization curve in Fig. 2).

In the perpendicular scattering geometry, a vertically elongated scattering pattern can be explained from having (at least) two possible origins. The first one is a strong uniaxial magnetic anisotropy (texture) with the easy axis along the horizontal in-plane direction. To investigate this possibility, we performed a series of field-dependent SANS measurements in the parallel scattering geometry. Here, it is important to remember that for a statistically-isotropic material, the parallel SANS cross section [Eq. (11)] is expected to be isotropic ($\theta$-independent) for symmetry reasons. Figure 6 displays the 2D magnetic SANS cross section of HPT-Ni in the parallel geometry. The magnetic SANS patterns are isotropic at all $H$ investigated, which suggests that no significant uniaxial in-plane magnetic texture exists in HPT-Ni, which might be expected due to the HPT treatment. Therefore, this possibility is excluded as the origin of the vertical elongation of the SANS pattern seen in the perpendicular geometry.

The other possible reason for a vertically elongated scattering pattern resides in the third term in Eq. (3) with a $\sin^2\theta$ dependence. In the approach-to-saturation regime, where the spin misalignment from the direction of the applied field is small, the micromagnetic SANS theory rigorously predicts $\Delta \widetilde{M}_z^2 = 0$ up to the second order of the small spin-misalignment amplitude. [Note that first-order in $\widetilde{\mathbf{M}}$ corresponds to a second-order contribution in $d\Sigma_{SM}/d\Omega$.] It is, however, entirely possible that in highly inhomogeneous materials the higher-order terms become non-negligible, breaking this property. This was already explored theoretically in [22] and experimentally in [40] by considering the following combination of cross section values at $\theta = 0°$ ($d\Sigma/d\Omega_{SM,0°}$) and at $\theta = 90°$ ($d\Sigma/d\Omega_{SM,90°}$) measured in the perpendicular geometry:

$$\Delta\Sigma_{SM}(q,H) = \frac{d\Sigma}{d\Omega_{SM,0°}}(q,H) - 2\frac{d\Sigma}{d\Omega_{SM,90°}}(q,H), \qquad (13)$$

which is identically zero in the second-order SANS theory [compare Eqs. (3) – (10)] and contains a leading third-order contribution in the amplitude of spin misalignment. Figures 7(a) and (b) show $d\Sigma/d\Omega_{SM,0°}$ and $d\Sigma/d\Omega_{SM,90°}$ of HPT-Ni at applied fields between 0.1 T and 4 T. Similarly to the corresponding $2\pi$-radially-averaged spin-misalignment SANS cross sections themselves (Fig. 5), the intensities drastically decrease with increasing $H$ for both $\theta$ directions. Figure 7(c) shows $\Delta\Sigma_{SM}$ computed using Eq. (13). Its magnitude decreases with increasing $H$ consistently with the reduction of the average spin misalignment. The values of $\Delta\Sigma_{SM}$ are of the order of $10^3$ cm$^{-1}$ at 0.1 T and $q = 0.04$ nm$^{-1}$, which is similar to the previous experimental results on NANOPERM as well as to the micromagnetic simulations of nanoporous Fe [40]. However, $\Delta\Sigma_{SM}$ in HPT-Ni turns out to be negative in the whole range of the applied fields between 0.1 T and 4 T. Theory [22] predicts that the third-order contribution to $\Delta\Sigma_{SM}$ may have negative values at some intermediate $q$-range. The existence of such a range is corroborated by the experimental study of NANOPERM [40], while in the micromagnetic simulations on nanoporous Fe, $\Delta\Sigma_{SM}$ at 0.6 T is positive for $q < 0.6$ nm$^{-1}$. Further extension of the third-order effect theory is therefore needed to fully explain the negative values of $\Delta\Sigma_{SM}$ for all $q$ as observed in HPT-Ni, in other words, the strong vertical elongation of the cross section so that its values at $\theta = 90°$ consistently exceed the values at $\theta = 0°$.

The angular ($\theta$) dependences of the scattering intensities for two selected fields, 0.2 T (far from the saturation regime) and 1.0 T (within the saturation regime), are shown in Figs. 8(a) and (b), respectively. The scattering profiles exhibit two peaks centered at $\theta = 90°$ and $\theta = 270°$. This agrees with the characteristics of a sine squared function. However, with increasing $q$ and $H$, additional shoulders appear on both sides of the peaks. Further, although the profiles have minima at $\theta = 0°$ and 180°, the intensities at the minima are non-zero. The relative intensities at the minima compared to the intensities at the peaks tend to decrease at higher fields. These features indicate that the profiles are different from just a simple sine squared behavior. The shoulders are probably attributed to the clover-leaf contribution, while the intensities at $\theta = 0°$ and 180° resemble the flying-saucer contribution. Hence, the magnetic scattering patterns of HPT-Ni are very likely composed of the clover-leaf and flying-saucer contributions as well as the $\sin^2\theta$ contribution.

To quantify these contributions to Eq. (3) we have fitted the magnetic cross sections using the function $S_H(\mathbf{q})R_{H,\perp}(q,\theta,H) + S_M(\mathbf{q})R_M(q,\theta,H) + S_{\sin}(q,H)\sin^2\theta$ with fixed values of $A = 8.2$ pJ/m [19] and $\mu_0 M_s = 0.6$ T. The fits, conducted for the azimuthal-angle ($\theta$) dependences from $q = 0.06$ nm$^{-1}$ to 0.2 nm$^{-1}$ at fields between 0.1 T and 1 T, are displayed as solid lines in Fig. 8 and are in good agreement with the data. This confirms that the presence of the $\sin^2\theta$ term can describe the vertical elongation of the observed cross sections well. Figure 9(a) shows the resultant $S_{\sin}(q,H)$ curves at fixed values of the magnetic field. Their functional $q$-dependencies are similar to the corresponding spin-misalignment scattering cross sections shown in Fig. 5. This means that $S_{\sin}(q,H)$ represents the dominant contribution to the spin-misalignment scattering cross section. Figure 9(b) depicts the magnetic field dependences of $S_{\sin}(q,H)$ at selected $q$-values. These reveal a plateau in the low-field region which transitions to a power-law behavior in the high-field region. The slope of the power law is -1.6 at $q = 0.06$ nm$^{-1}$. The transition point between the plateau and the power-law region shifts to higher field with increasing $q$.

At the moment, micromagnetic SANS theory can provide analytical solutions of the third-order effects for a few cases such as the contribution to the cross section along the $\theta = 0°$ and $90°$ directions, sufficient to compute $\Delta\Sigma_{SM}$, but not the complete angular dependence [22]. However, since the $S_{\sin}(q,H)$ contribution to Eq. (3) is related to $\Delta\widetilde{M}_z^2(\mathbf{q},H)$, its field dependence can be qualitatively interpreted using the law of approach to ferromagnetic saturation (LAFS) [41-46]. The LAFS describes the approach of the average $\widetilde{M}_z(\mathbf{q},H)|_{q=0}$ to the average saturation magnetization $\widetilde{M}_s$ as the field increases which, at large fields $H \to \infty$, is usually represented as a power law $\widetilde{M}_s - \widetilde{M}_z(\mathbf{q},H)|_{q=0} \sim H^{-n}$ or $\Delta\widetilde{M}_z^2(\mathbf{q},H)|_{q=0} \sim H^{-2n}$, with $n = 1/2$, 1, 3/2, 2, and 3 [41-46]. Several studies have also reported the transition to a plateau region in the low-field part for $H < M_s$ [42-44]. The field dependence of $S_{\sin}(q,H)$ is consistent with those previous studies. Thus, the observed exponent of -1.6 [Fig. 9(b)] can be explained by a LAFS with $H^{-0.8}$ and suggests that the misaligned spins in HPT-Ni are related to contributions with $n = 1/2$ and/or 1, which corresponds to the LAFS of amorphous ferromagnets, spherical defects, and dislocation dipoles [46]. The shift of the transition field may result from the $q$-dependence of the saturation magnetization.

A non-zero value of $S_{\sin}$ and the negative sign of $\Delta\Sigma_{SM}$ may be attributed to differences in the microstructures and magnetic properties between HPT-Ni and the other investigated systems, which break the assumptions of the original micromagnetic SANS theory both in the second and in the third order. One possibility is the large amplitude of the variations in the material constants such as $M_s$ and $A$, which amplify the errors of the Taylor series expansion in these theories. The shape of the inclusions (defects) also plays an important role. Previous studies pointed out that the third-order contribution is larger for layer-like inhomogeneities compared to spherical ones [32,40]. In HPT-Ni, spin misalignments probably originate from the grain-boundary phase as determined from the magnetic measurements and other possible defects [26], which generally have a two-dimensional structure and no particular shape, respectively. By contrast, the inclusions are approximately spherical both in NANOPERM and in nanoporous Fe [40]. Hence, the two-dimensional nature of the grain-boundary phase in HPT-Ni could be the reason for the enhancement of the higher-order contribution to the cross section with a characteristic $\sin^2\theta$ angular dependence. Since the cross-section along the $\theta = 90°$ direction is subtracted in Eq. (13), this makes the whole $\Delta\Sigma_{SM}$ negative.

## V. CONCLUSION

We have performed magnetic-field-dependent unpolarized SANS experiments to investigate the role played by artificially-created crystal defects on the spin texture of HPT-Ni. The analysis of the field-dependent magnetic SANS data reveals that the magnetic neutron scattering cross section is dominated by the spin-misalignment contribution. Whereas the standard micromagnetic SANS theory in the second order predicts clover-leaf-type and flying-saucer-type scattering patterns, the spin-misalignment scattering in HPT-Ni exhibits a predominant $\sin^2\theta$ contribution, which remains visible in applied fields up to 4 T. The presence of this contribution cannot be explained by second-order micromagnetic SANS theory, while its strength relatively to the rest of the cross section is also beyond of what can be explained by third-order micromagnetic SANS theory [22]. Just like in the approach-to-saturation theory [32], the reason behind the significance of higher-order effects lies, probably, in the presence of a high-density of high-amplitude fluctuations of the material parameters induced via HPT straining. We hope that our experimental results will fuel the further development of the theory of higher-order magnetic SANS.

**ACKNOWLEDGMENTS**

The authors acknowledge the Australian Nuclear Science and Technology Organisation (ANSTO), Lucas Heights, Australia (proposal No. P5264), and the Institut Laue-Langevin (ILL), Grenoble, France, for the provision of neutron beamtime. The authors are grateful to Mr. K. Yamamoto, Dr. N. Sato, the Sample Environment groups, particularly Dr. Norman Booth (ANSTO) and Dr. Robert Cubitt (ILL) for their help in the experiment. K. L. Metlov acknowledges the support of the Russian Science Foundation under the project RSF 21-11-00325. This work was financially supported by KAKENHI Grant Number 19K05102, Japan Science and Technology Agency (JST) under Collaborative Research Based on Industrial Demand "Heterogeneous Structure Control: Towards Innovative Development of Metallic Structural Materials" (Grant No. JPMJSK1511), the General User Program for Neutron Scattering Experiments, Institute for Solid State Physics, the University of Tokyo (proposal Nos. 15563, 16554, and 16566).

**REFERENCES**

[1] R. Z. Valiev, A. P. Zhilyaev, and T. G. Langdon, *Bulk Nanostructured Materials: Fundamentals and Applications* (John Wiley & Sons, Somerset, 2013).

[2] R. Z. Valiev, Y. Estrin, Z. Horita, T. G. Langdon, M. J. Zehetbauer, and Y. Zhu, JOM **68**, 1216 (2016).

[3] M. Tane, Y. Okuda, Y. Todaka, H. Ogi, and A. Nagakubo, Acta Mater. **61**, 7543 (2013).

[4] A. P. Zhilyaev and T. G. Langdon, Prog. Mater. Sci. **53**, 893 (2008).

[5] K. Edalati and Z. Horita, Mater. Sci. Eng. A **652**, 325 (2016).

[6] W. Li, L. Li, Y. Nan, Z. Xu, X. Zhang, A. G. Popov, D. V. Gunderov, and V. V. Stolyarov, J. Appl. Phys. **104**, 023912 (2008).

[7] S. Lee, K. Edalati, H. Iwaoka, Z. Horita, T. Ohtsuki, and T. Ohkochi, Philos. Mag. Lett. **94**, 639 (2014).

[8] C. M. Cepeda-Jiménez, A. Hernando, J. M. Barandiarán, M. T. Pérez-Prado, Scr. Mater. **118**, 41 (2016).

[9] C. M. Cepeda-Jiménez, J. I. Beltrán, A. Hernando, M. A. García, F. Ynduráin, A. Zhilyaev, M. T. Pérez-Prado, Acta Mater. **123**, 206 (2017).

[10] J. B. Goodenough, Phys. Rev. **95**, 917 (1954).

[11] J. Degauque, B. Astie, J. L. Porteseil, and R. Vergne, J. Magn. Magn. Mater. **26**, 261 (1982).

[12] G. Herzer, J. Magn. Magn. Mater. **294**, 99 (2005).

[13] K. Y. Mulyukov, G. F. Korznikova, R. Z. Abdulov, and R. Z. Valiev, J. Magn. Magn. Mater. **89**, 207 (1990).

[14] K. Y. Mulyukov, S. B. Khaphizov, and R. Z. Valiev, Phys. Stat. Sol. (a) **133**, 447 (1992).

[15] G. F. Korznikova, K. Y. Mulyukov, V. N. Timofeyev, and R. Z. Valiev, J. Magn. Magn. Mater. **135**, 46 (1994).

[16] G. F. Korznikova, J. Microsc. **239**, 239 (2010).

[17] S. Scheriau, M. Kriegisch, S. Kleber, N. Mehboob, R. Grössinger, and R. Pippan, J. Magn. Magn. Mater. **322**, 2984 (2010).

[18] J. M. Rojo, A. Hernando, M. El Ghannami, A. García-Escorial, M. A. González, R. García-Martínez, and L. Ricciarelli, Phys. Rev. Lett. **76**, 4833 (1996).

[19] J. Weissmüller, A. Michels, J. G. Barker, A. Wiedenmann, U. Erb, and R. D. Shull, Phys. Rev. B **63**, 214414 (2001).

[20] A. Wiedenmann, J. Appl. Crystallogr. **33**, 428 (2000).

[21] A. Michels, J. Phys.: Cond. Mat. **26**, 383201 (2014).

[22] K. L. Metlov and A. Michels, Phys. Rev. B **91**, 054404 (2015).

[23] D. Mettus and A. Michels, J. Appl. Crystallogr. **48**, 1437 (2015).

[24] S. Mühlbauer, D. Honecker, É. A. Périgo, F. Bergner, S. Disch, A. Heinemann, S. Erokhin, D. Berkov, C. Leighton, M. R. Eskildsen, and A. Michels, Rev. Mod. Phys. **91**, 015004 (2019).

[25] Y. Oba, N. Adachi, Y. Todaka, E. P. Gilbert, and H. Mamiya, Phys. Rev. Research **2**, 033473 (2020).

[26] M. Bersweiler, E. Pratami Sinaga, I. Peral, N. Adachi, P. Bender, N.-J. Steinke, E.P. Gilbert, Y. Todaka, A. Michels, and Y. Oba, Phys. Rev. Mater. **5**, 044409 (2021).

[27] K. Wood, J. P. Mata, C. J. Garvey, C.-M. Wu, W. A. Hamilton, P. Abbeywick, D. Bartlett, F. Bartsch, P. Baxter, N. Booth, W. Brown, J. Christoforidis, D. Clowes, T. d'Adam, F. Darmann, M. Deura, S. Harrison, N. Hauser, G. Horton, D. Federici, F. Franceschini, P. Hanson, E. Imamovic, P. Imperia, M. Jones, S. Kennedy, S. Kim, T. Lam, W. T. Lee, M. Lesha, D. Mannicke, T. Noakes, S. R. Olsen, J. C. Osborn, D. Penny, M. Perry, S. A. Pullen, R. A. Robinson, J. C. Schulz, N. Xiong, and E. P. Gilbert, J. Appl. Crystallogr. **51**, 294 (2018).

[28] S. R. Kline, J. Appl. Crystallogr. **39**, 895 (2006).

[29] C. D. Dewhurst, I. Grillo, D. Honecker, M. Bonnaud, M. Jacques, C. Amrouni, A. Perillo-Marcone, G. Manzin, and R. Cubitt, J. Appl. Crystallogr. **49**,14 (2016).

[30] C. D. Dewhurst, Graphical Reduction and Analysis SANS Program for Matlab$^{TM}$ (Institut Laue–Langevin, Grenoble, 2018), https://www.ill.eu/users/support-labs-infrastructure/software-scientific-tools/grasp.

[31] E. Schlömann, J. Appl. Phys. **38**, 5027 (1967).

[32] E. Schlömann, J. Appl. Phys. **42**, 5798 (1971).

[33] A. T. Aldred, Phys. Rev. B **11**, 2597 (1975).

[34] J.-P. Bick, D. Honecker, F. Döbrich, K. Suzuki, E. P. Gilbert, H. Frielinghaus, J. Kohlbrecher, J. Gavilano, E. M. Forgan, R. Schweins, P. Lindner, R. Birringer, and A. Michels, Appl. Phys. Lett. **102**, 022415 (2013).

[35] M. Bersweiler, P. Bender, L. G. Vivas, M. Albino, M. Petrecca, S. Mühlbauer, S. Erokhin, D. Berkov, C. Sangregorio, and A. Michels, Phys. Rev. B **100**, 144434 (2019).

[36] J. Weissmüller, A. Michels, D. Michels, A. Wiedenmann, C. E. Krill III, H. M. Sauer, and R. Birringer, Phys. Rev. B **69**, 054402 (2004).

[37] D. Honecker, C. D. Dewhurst, K. Suzuki, S. Erokhin, and A. Michels, Phys. Rev. B **88**, 094428 (2013).

[38] É. A. Périgo, E. P. Gilbert, K. L. Metlov, and A. Michels, New J. Phys. **16**, 123031 (2014).

[39] F. Döbrich, J. Kohlbrecher, M. Sharp, H. Eckerlebe, R. Birringer, and A. Michels, Phys. Rev. B **85**, 094411 (2012).

[40] K. L. Metlov, K. Suzuki, D. Honecker, and A. Michels, Phys. Rev. B **101**, 214410 (2020).

[41] W. F. Brown, Phys. Rev. **58**, 736 (1940).

[42] W. F. Brown, Phys. Rev. **60**, 139 (1941).

[43] M. Fähnle and H. Kronmüller, J. Magn. Magn. Mater. **8**, 149 (1978).

[44] H. Kronmüller, IEEE Trans. Magn. MAG-**15**, 1218 (1979).

[45] E. M. Chudnovsky, W. M. Saslow, and R. A. Serota, Phys. Rev. B **33**, 251 (1986).

[46] H. Kronmüller, General Micromagnetic Theory, in *Handbook of Magnetism and Advanced Magnetic Materials*, edited by H. Kronmüller, S. Parkin, J. E. Miltat, and M. R. Scheinfein, (John Wiley & Sons, New York, 2007), pp. 1-39.

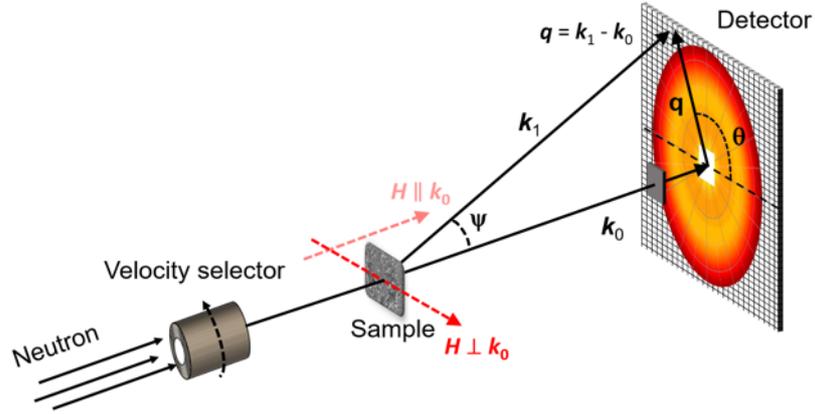

FIG. 1 Sketch of the small-angle neutron scattering geometry. The wave vectors of the incoming and scattered neutrons are, respectively, $\mathbf{k}_0$ and $\mathbf{k}_1$, whereas $\mathbf{q}$ denotes the momentum-transfer vector. The applied magnetic field $\mathbf{H}$ is perpendicular or parallel to the incoming neutron beam, and the angle $\theta$ specifies the angular anisotropy of the scattering pattern on the two-dimensional detector.

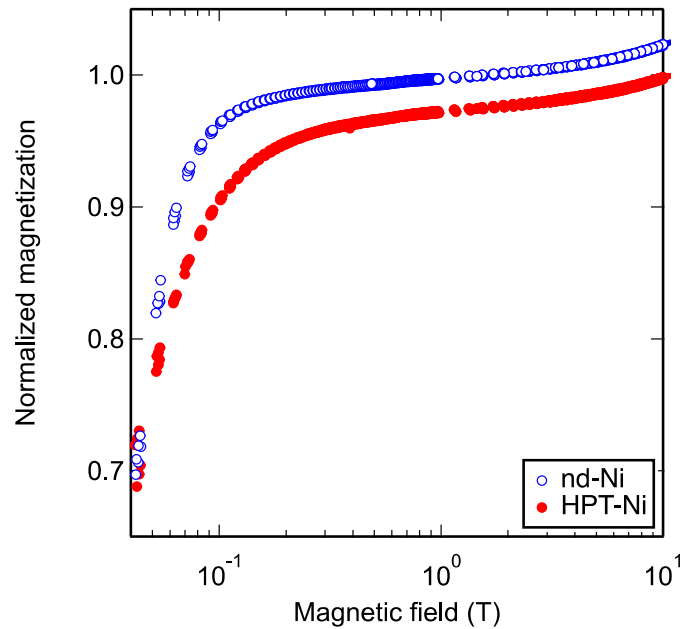

FIG. 2 Room-temperature magnetization curves of HPT-Ni and nd-Ni normalized by the saturation magnetization of bulk Ni ($M_s$ = 55.5 Am$^2$/kg [33]) (only the upper right quadrant is shown). Filled and open circles denote, respectively, the magnetizations of HPT-Ni and nd-Ni (semi-log scale).

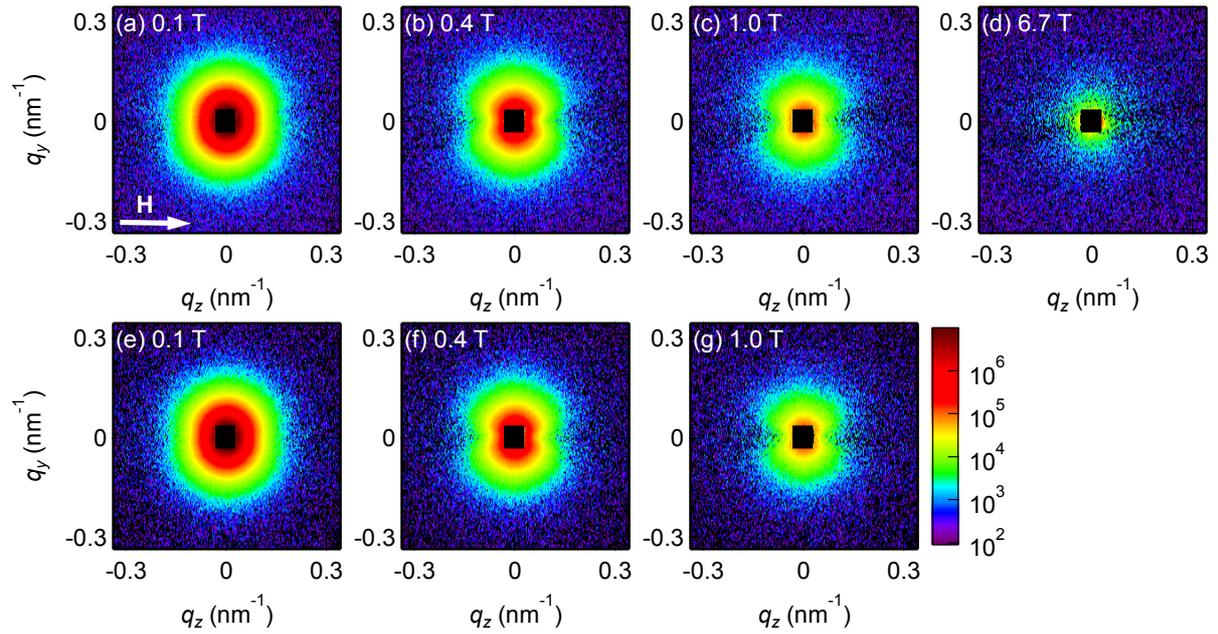

FIG. 3 Top panel: two-dimensional (2D) total (nuclear + magnetic) SANS cross section of HPT-Ni measured at (a) 0.1 T, (b) 0.4 T, (c) 1.0 T, and (d) 6.7 T. The magnetic field is applied horizontally in the detector plane and perpendicular to the incoming neutron beam (perpendicular scattering geometry). Bottom panel: corresponding purely 2D magnetic SANS cross section determined at (e) 0.1 T, (f) 0.4 T, and (g) 1.0 T.

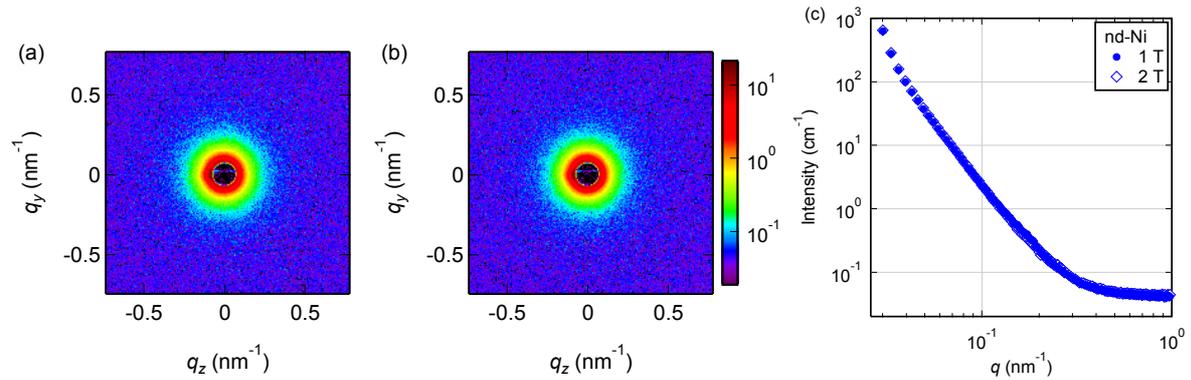

FIG. 4 Two-dimensional total (nuclear + magnetic) SANS cross sections of non-deformed Ni (nd-Ni) at the selected fields of (a) 1 T and at (b) 2 T (applied field is perpendicular to the incident neutron beam direction). Panel (c) displays the 2π-radially-averaged scattering profiles (log-log plot). Filled circles and open diamonds denote the profiles at 1 T and 2 T, respectively.

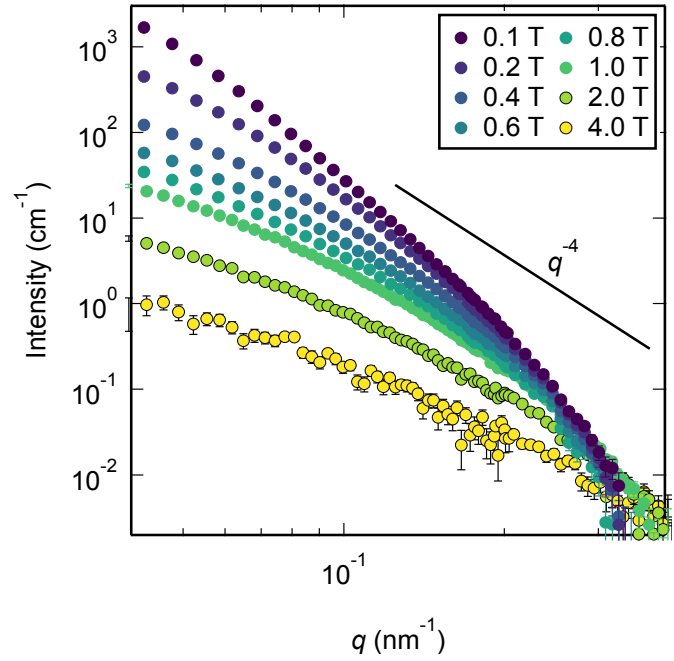

FIG. 5 Field dependence of the 2π-radially-averaged spin-misalignment scattering cross section. Field values are specified in the inset (log-log plot). Solid line is a visual guide for a $q^{-4}$-dependency.

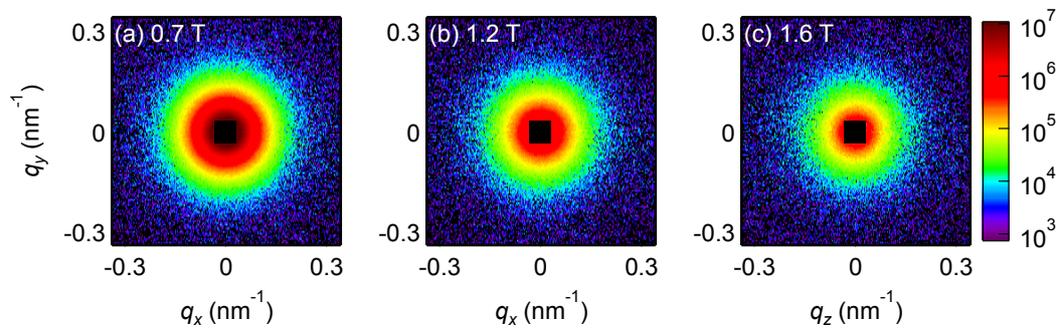

FIG. 6 Two-dimensional magnetic SANS cross section of HPT Ni at (a) 0.7 T, (b) 1.2 T, and (c) 1.6 T. The total nuclear and magnetic SANS signal at 6.7 T has been subtracted. The magnetic field is applied parallel to the neutron beam and is normal to the detector plane (parallel scattering geometry).

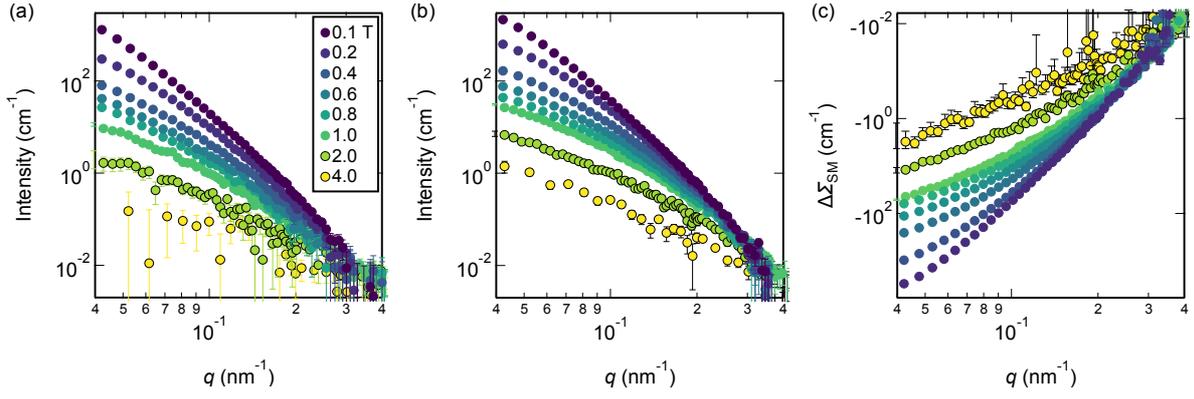

FIG. 7 Field-dependence of the spin-misalignment scattering cross section (a) parallel ($\theta = 0°$) and (b) perpendicular ($\theta = 90°$) to the applied magnetic field. Field values (in T) are specified in the inset. (c) $\Delta\Sigma_{SM}$ evaluated according to Eq. (15) from the parallel and perpendicular scattering profiles (log-log plots).

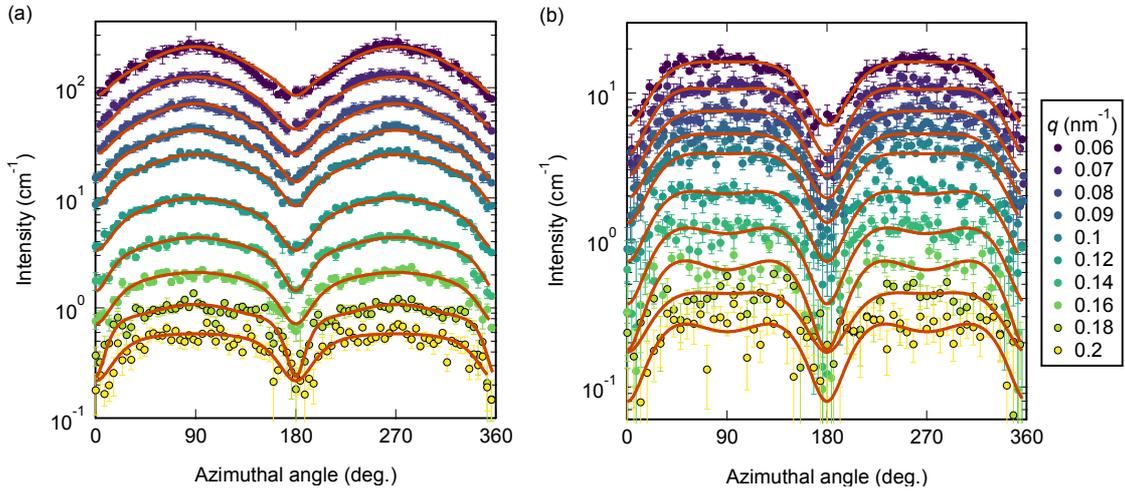

FIG. 8 Azimuthal-angle dependence of magnetic scattering intensities at the selected fields of (a) 0.2 T (far from the saturation regime) and at (b) 1.0 T (within the saturation regime). Shown is the magnetic SANS cross section (highest-field data subtracted). Values of $q$ (in nm$^{-1}$) increase from top to bottom (see inset). Solid lines: fit curves using the function $S_H(\mathbf{q})R_{H,\perp}(q,\theta,H) + S_M(\mathbf{q})R_M(q,\theta,H) + S_{sin}(q,H)\sin^2\theta$ (semi-log plot).

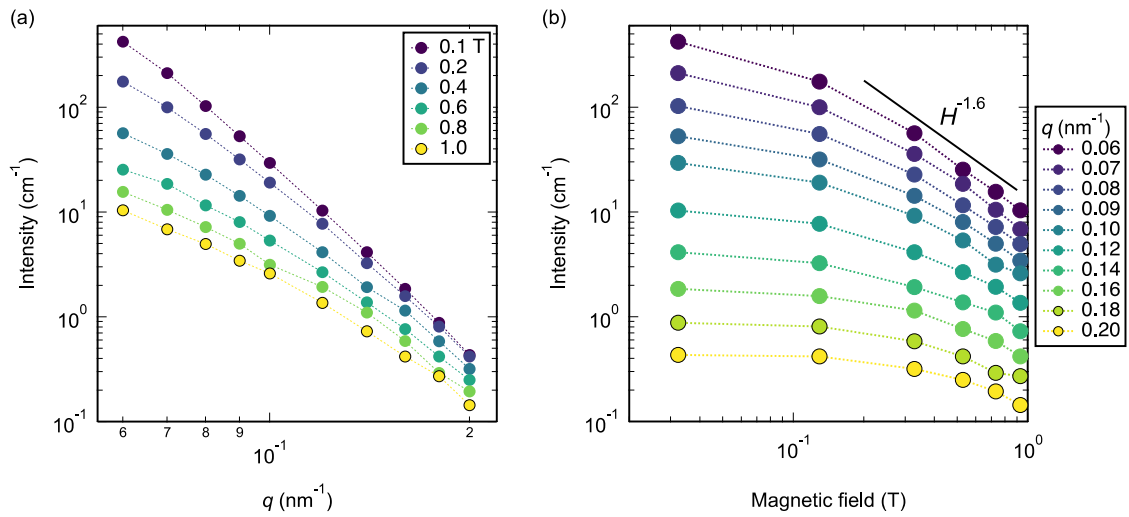

FIG. 9 (a) Momentum-transfer dependence of the $S_{\sin}(q,H)$ contribution at magnetic fields of 0.1, 0.2, 0.4, 0.6, 0.8, and 1 T (from top to bottom). (b) Field dependence of $S_{\sin}(q,H)$ at $q$ = 0.06, 0.07, 0.08, 0.09, 0.10, 0.12, 0.14, 0.16, 0.18, and 0.20 nm$^{-1}$ (from top to bottom). Solid line is a visual guide for $H^{-1.6}$ (log-log scales).